\documentclass[10pt]{iopart}

\usepackage{graphicx}
\usepackage{dcolumn}
\usepackage{bm}
\usepackage{color}

\begin{document}

\title[EIC stabilization]{Theoretical analysis of energetic-ion-driven resistive interchange mode stabilization strategies using a Landau closure model}


\author{J. Varela}
\ead{jacobo.varela@nifs.ac.jp}
\address{National Institute for Fusion Science, National Institute of Natural Science, Toki, 509-5292, Japan}
\author{S. Ohdachi}
\address{National Institute for Fusion Science, National Institute of Natural Science, Toki, 509-5292, Japan}
\author{K. Y. Watanabe}
\address{National Institute for Fusion Science, National Institute of Natural Science, Toki, 509-5292, Japan}
\author{D. A. Spong}
\address{Oak Ridge National Laboratory, Oak Ridge, Tennessee 37831-8071, USA}
\author{L. Garcia}
\address{Universidad Carlos III de Madrid, 28911 Leganes, Madrid, Spain}
\author{R. Seki}
\address{SOKENDAI, Department of Fusion Science, Toki/Gifu and National Institute for Fusion Science, National Institute of Natural Science, Toki, 509-5292, Japan}

\date{\today}

\begin{abstract}
The aim of the present study is to perform a theoretical analysis of different strategies to stabilize energetic-ion-driven resistive interchange mode (EIC) in LHD plasma. We use a reduced MHD for the thermal plasma coupled with a gyrofluid model for the energetic particles (EP) species. The hellically trapped EP component is introduced through a modification of the drift frequency to include their precessional drift. The stabilization trends of the $1/1$ EIC observed experimentally with respect to the thermal plasma density and temperature are reproduced by the simulations, showing a reasonable agreement with the data. The LHD operation scenarios with stable $1/1$ EIC are identified, leading to the stabilization of the $1/1$ EIC if the thermal plasma density and temperature are above a given threshold. The $1/1$ EIC are also stabilized if the rotational transform is modified in a way that the $1/1$ rational surface is located further away than $0.9$ times the normalized radius, or the magnetic shear in the plasma periphery is enhanced. Also, LHD discharges with large magnetic fields show a higher EIC destabilization threshold with respect to the thermal plasma density. If the perpendicular NBI deposition region is moved further inward than $0.875$ times the normalized radius the $1/1$ EIC are also stabilized. In addition, increasing the perpendicular NBI voltage such that the EP energy is higher than $30$ keV stabilizes the $1/1$ EIC. Moreover, Deuterium plasmas show a higher stability threshold for the $1/1$ EIC than Hydrogen plasmas. The experimental data shows a larger time interval between EIC events as the power of the tangential NBI is increased providing that the perpendicular NBI power is at least $13$ MW. This implies a stabilizing effect of the tangential NBI.
\end{abstract}

%
%
%
%
%


\vspace{2pc}

\maketitle

\ioptwocol

\section{Introduction \label{sec:introduction}}

The $1/1$ energetic-ion-driven resistive interchange mode (EIC) are observed in Large Helical Device (LHD) plasma with unstable resistive interchange modes (RIC), destabilized in the magnetic hill region at the plasma periphery \cite{1,2,3,4,5}. The $1/1$ EIC is a bursting instability triggered if the perpendicular NBI injection overcomes some threshold, chirping down from $9$ to $4$ kHz before stabilization \cite{6,7,8}. The precessional motion of the helically trapped EP generated by the perpendicular NBI resonates with the RIC causing EP losses \cite{9,10}. Consequently, the EIC events can be grouped in the family of the energetic particle modes (EPM) \cite{11}, similar to the fish-bones oscillations \cite{7,12}. 

The transport of fusion produced alpha particles, energetic hydrogen neutral beams and ion cyclotron resonance heated particles (ICRF) can be enhanced by energetic particle driven instabilities \cite{13,14,15}, leading to a decrease of the heating efficiency in helical devices such as LHD and W7-AS stellarators or tokamaks such as JET and DIII-D \cite{16,17,18,19,20,21}. The EP losses are enhanced because there is a resonance between the unstable mode frequency and the EP drift, bounce or transit frequencies. In particular, the EPM are unstable for frequencies in the shear Alfven continua if the continuum damping is not strong enough to stabilize them \cite{22,23,24,25,26}. 

LHD is a helical device heated by three NBI lines almost parallel to the magnetic axis with an energy of 180 keV and two NBI perpendicular to the magnetic axis with an energy of 32 keV. The $1/1$ EIC are destabilized in discharges with strong perpendicular NBI injection and low thermal ion density, both in Hydrogen and Deuterium plasma \cite{27}, limiting the device performance.

The stability of the RIC was widely analyzed theoretically and experimentally by other authors, although the effect of the EIC on the LHD performance is a new an important topic to study, because high $\beta$ LHD discharges are strongly limited by this instability leading to an inefficient plasma heating. Several stabilization trends to reduce or mitigate the $1/1$ EIC were identified experimentally, for example increasing the thermal plasma density and the thermal plasma temperature above a given threshold by the application of electron cyclotron heating (ECH) \cite{28,29}, or by the application of resonant magnetic perturbations (RMP) \cite{30}. The aim of the present study is to perform a theoretical analysis of the $1/1$ EIC stability in different LHD operational scenarios. Optimization trends will be identified to avoid triggering the $1/1$ EIC. These optimizations will involve the thermal plasma parameters, the operational regime of the perpendicular NBI, the magnetic field topology and intensity or the thermal plasma and perpendicular NBI species. First, a parametric study is performed for a range of thermal plasma density and temperature values, identifying the LHD operational scenarios with stable $1/1$ EIC and comparing the simulation results with the experimental trends. Next, new optimization trends with respect to rotational transform and fast ion drive are analyzed. In addition, the $1/1$ EIC stability with respect to the perpendicular NBI voltage and deposition region is analyzed. Lastly, the effect of the thermal plasma and perpendicular NBI species are also studied.

The simulations are performed using the FAR3D code \cite{31,32,33}. The numerical model solves the reduced linear resistive MHD equations and the moment equations of the energetic ion density and parallel velocity \cite{34,35}, including for the appropriate Landau closure relations the linear wave-particle resonance effects required for Landau damping/growth, as well as the parallel momentum response of the thermal plasma required for coupling to the geodesic acoustic waves \cite{36}. Six field variables evolves starting from an equilibria calculated by the VMEC code \cite{37}.

This paper is organized as follows. The model equations, numerical scheme and equilibrium properties are described in section \ref{sec:model}. The $1/1$ EIC stabilization strategies analyzed experimentally are reproduced in section \ref{sec:exp}. The new optimization strategies with respect to the rotational transform and fast ion (EP) drive are studied in section \ref{sec:Bfield}. $1/1$ EIC stabilization strategies linked to the perpendicular NBI operational regime are analyzed in section \ref{sec:EP}. Next, a comparison of the $1/1$ EIC stability between Hydrogen/Deuterium plasma heated by an Hydrogen/Deuterium NBI is shown in section \ref{sec:Species}. The effect of the multiple EP species is also analyzed in section \ref{sec:multiple}. Finally, the conclusions of this paper are presented in section \ref{sec:conclusions}.

\section{Equations and numerical scheme \label{sec:model}}

Following the method employed in Ref.\cite{38}, a reduced set of equations for high-aspect ratio configurations and moderate $\beta$-values (of the order of the inverse aspect ratio) is derived retaining the toroidal angle variation based upon an exact three-dimensional equilibrium that assumes closed nested flux surfaces. The effect of the energetic particle population in the plasma stability is included through moments of the fast ion kinetic equation truncated with a closure relation \cite{39}, describing the evolution of the energetic particle density ($n_{f}$) and velocity moments parallel to the magnetic field lines ($v_{||f}$). The coefficients of the closure relation are selected to match analytic TAE growth rates based upon a two-pole approximation of the plasma dispersion function.

The model formulation assumes high aspect ratio, medium $\beta$ (of the order of the inverse aspect ratio $\varepsilon=a/R_0$), small variation of the fields and small resistivity. The plasma velocity and perturbation of the magnetic field are defined as
\begin{equation}
 \mathbf{v} = \sqrt{g} R_0 \nabla \zeta \times \nabla \Phi, \quad\quad\quad  \mathbf{B} = R_0 \nabla \zeta \times \nabla \psi,
\end{equation}
where $\zeta$ is the toroidal angle, $\Phi$ is a stream function proportional to the electrostatic potential, and $\tilde \psi$ is the perturbation of the poloidal flux.

The equations, in dimensionless form, are
\begin{equation}
\frac{\partial \tilde \psi}{\partial t} =  \sqrt{g} B \nabla_\| \Phi  + \eta \varepsilon^2 J \tilde J^\zeta
\end{equation}
\begin{eqnarray} 
\frac{{\partial \tilde U}}{{\partial t}} =  - v_{\zeta,eq} \frac{\partial \tilde U}{\partial \zeta} \nonumber\\
+ \sqrt{g} B  \nabla_{||} \tilde J^{\zeta} - \frac{1}{\rho} \left( \frac{\partial J_{eq}}{\partial \rho} \frac{\partial \tilde \psi}{\partial \theta} - \frac{\partial J_{eq}}{\partial \theta} \frac{\partial \tilde \psi}{\partial \rho}    \right) \nonumber\\
- {\frac{\beta_0}{2\varepsilon^2} \sqrt{g} \left( \nabla \sqrt{g} \times \nabla \tilde p \right)^\zeta } -  {\frac{\beta_f}{2\varepsilon^2} \sqrt{g} \left( \nabla \sqrt{g} \times \nabla \tilde n_f \right)^\zeta }
\end{eqnarray} 
\begin{eqnarray}
\label{pressure}
\frac{\partial \tilde p}{\partial t} = - v_{\zeta,eq} \frac{\partial \tilde p}{\partial \zeta} + \frac{dp_{eq}}{d\rho}\frac{1}{\rho}\frac{\partial \tilde \Phi}{\partial \theta} \nonumber\\
 +  \Gamma p_{eq}  \left[{\left( \nabla \sqrt{g} \times \nabla \tilde \Phi \right)^\zeta - \nabla_\|  \tilde v_{\| th} }\right] 
\end{eqnarray} 
\begin{eqnarray}
\label{velthermal}
\frac{{\partial \tilde v_{\| th}}}{{\partial t}} = - v_{\zeta,eq} \frac{\partial \tilde v_{||th}}{\partial \zeta} -  \frac{\beta_0}{2n_{0,th}} \nabla_\| p 
\end{eqnarray}
\begin{eqnarray}
\label{nfast}
\frac{{\partial \tilde n_f}}{{\partial t}} = - v_{\zeta,eq} \frac{\partial \tilde n_{f}}{\partial \zeta} - \frac{v_{th,f}^2}{\varepsilon^2 \omega_{cy}}\ \Omega_d (\tilde n_f) - n_{f0} \nabla_\| \tilde v_{\| f}   \nonumber\\
-  n_{f0} \, \Omega_d (\tilde \Phi) + n_{f0} \, \Omega_* (\tilde  \Phi) 
\end{eqnarray}
\begin{eqnarray}
\label{vfast}
\frac{{\partial \tilde v_{\| f}}}{{\partial t}} = - v_{\zeta,eq} \frac{\partial \tilde v_{||f}}{\partial \zeta}  -  \frac{v_{th,f}^2}{\varepsilon^2 \omega_{cy}} \, \Omega_d (\tilde v_{\| f}) \nonumber\\
- \left( \frac{\pi}{2} \right)^{1/2} v_{th,f} \left| \nabla_\| \tilde v_{\| f}  \right| - \frac{v_{th,f}^2}{n_{f0}} \nabla_\| n_f + v_{th,f}^2 \, \Omega_* (\tilde \psi) 
\end{eqnarray}
Equation (2) is derived from Ohm’s law coupled with Faraday’s law, equation (3) is obtained from the toroidal component of the momentum balance equation after applying the operator $\nabla \wedge \sqrt{g}$, equation (4) is obtained from the thermal plasma continuity equation with compressibility effects and equation (5) is obtained from the parallel component of the momentum balance. Here, $U =  \sqrt g \left[{ \nabla  \times \left( {\rho _m \sqrt g {\bf{v}}} \right) }\right]^\zeta$ is the toroidal component of the vorticity, $\rho_m$ the ion and electron mass density, $\rho = \sqrt{\phi_{N}}$ the effective radius with $\phi_{N}$ the normalized toroidal flux and $\theta$ the poloidal angle. The perturbation of the toroidal current density $\tilde J^{\zeta}$ is defined as:
\begin{eqnarray}
\tilde J^{\zeta} =  \frac{1}{\rho}\frac{\partial}{\partial \rho} \left(-\frac{g_{\rho\theta}}{\sqrt{g}}\frac{\partial \tilde \psi}{\partial \theta} + \rho \frac{g_{\theta\theta}}{\sqrt{g}}\frac{\partial \tilde \psi}{\partial \rho} \right) \nonumber\\
- \frac{1}{\rho} \frac{\partial}{\partial \theta} \left( \frac{g_{\rho\rho}}{\sqrt{g}}\frac{1}{\rho}\frac{\partial \tilde \psi}{\partial \theta} + \rho \frac{g_{\rho \theta}}{\sqrt{g}}\frac{\partial \tilde \psi}{\partial \rho} \right)
\end{eqnarray}
$v_{||th}$ is the parallel velocity of the thermal particles and $v_{\zeta,eq}$ is the equilibrium toroidal rotation. $\beta_{0}$ is the equilibrium $\beta$ at the magnetic axis, $\beta_{f}$ is the maximum value of the EP $\beta$ (located at the magnetic axis in the on-axis cases but not in the off-axis cases) and $n_{f0}$ is the EP radial density profile normalized to the local maxima. $\Phi$ is normalized to $a^2B_{0}/\tau_{A0}$ and $\tilde\psi$ to $a^2B_{0}$ with $\tau_{A0}$ the Alfv\' en time $\tau_{A0} = R_0 (\mu_0 \rho_m)^{1/2} / B_0$. The radius $\rho$ is normalized to a minor radius $a$; the resistivity to $\eta_0$ (its value at the magnetic axis); the time to the Alfv\' en time; the magnetic field to $B_0$ (the averaged value at the magnetic axis); and the pressure to its equilibrium value at the magnetic axis. The Lundquist number $S$ parameter is the ratio of the resistive time $\tau_{R} = a^2 \mu_{0} / \eta_{0}$ to the Alfv\' en time. $\rlap{-} \iota$ is the rotational transform, $v_{th,f} = \sqrt{T_{f}/m_{f}}$ is the radial profile of the energetic particle thermal velocity normalized to the Alfv\' en velocity at the magnetic axis $v_{A0}$ and $\omega_{cy}$ the energetic particle cyclotron frequency normalized to $\tau_{A0}$. $q_{f}$ is the charge, $T_{f}$ is the radial profile of the effective EP temperature and $m_{f}$ is the mass of the EP. The $\Omega$ operators are defined as:
\begin{eqnarray}
\label{eq:omedrift}
\Omega_d = \frac{\epsilon^2 \pi \rho^2 \omega_{b}}{d_{b}} \left[ \frac{\partial}{\partial \theta} \left( \frac{1}{\sqrt{g}} \right) \right]^{-1} \cdot \nonumber\\
\Bigg\{ \frac{1}{2 B^4 \sqrt{g}}  \left[  \left( \frac{I}{\rho} \frac{\partial B^2}{\partial \zeta} - J \frac{1}{\rho} \frac{\partial B^2}{\partial \theta} \right) \frac{\partial}{\partial \rho}\right] \nonumber\\
-   \frac{1}{2 B^4 \sqrt{g}} \left[ \left( \rho \beta_* \frac{\partial B^2}{\partial \zeta} - J \frac{\partial B^2}{\partial \rho} \right) \frac{1}{\rho} \frac{\partial}{\partial \theta} \right] \nonumber\\ 
+ \frac{1}{2 B^4 \sqrt{g}} \left[ \left( \rho \beta_* \frac{1}{\rho} \frac{\partial B^2}{\partial \theta} -  \frac{I}{\rho} \frac{\partial B^2}{\partial \rho} \right) \frac{\partial}{\partial \zeta} \right] \Bigg\}
\end{eqnarray}

\begin{eqnarray}
\label{eq:omestar}
\Omega_* = \frac{1}{B^2 \sqrt{g}} \frac{1}{n_{f0}} \frac{d n_{f0}}{d \rho} \left( \frac{I}{\rho} \frac{\partial}{\partial \zeta} - J \frac{1}{\rho} \frac{\partial}{\partial \theta} \right).
\end{eqnarray}
Here the $\Omega_{d}$ operator models the average drift velocity of a helically trapped particle and $\Omega_{*}$ models the diamagnetic drift frequency. The parameter $\omega_{b}=100$ kHz indicates the bounce frequency and $d_{b}=0.01$ m the bounce length of the helically trapped EP guiding center. For more details regarding the derivation of the average drift velocity operator please see Ref.\cite{11}.

We also define the parallel gradient and curvature operators as
\begin{equation}
\label{eq:gradpar}
\nabla_\| f = \frac{1}{B \sqrt{g}} \left( \frac{\partial \tilde f}{\partial \zeta} +  \rlap{-} \iota \frac{\partial \tilde f}{\partial \theta} - \frac{\partial f_{eq}}{\partial \rho}  \frac{1}{\rho} \frac{\partial \tilde \psi}{\partial \theta} + \frac{1}{\rho} \frac{\partial f_{eq}}{\partial \theta} \frac{\partial \tilde \psi}{\partial \rho} \right)
\end{equation}
\begin{equation}
\label{eq:curv}
\sqrt{g} \left( \nabla \sqrt{g} \times \nabla \tilde f \right)^\zeta = \frac{\partial \sqrt{g} }{\partial \rho}  \frac{1}{\rho} \frac{\partial \tilde f}{\partial \theta} - \frac{1}{\rho} \frac{\partial \sqrt{g} }{\partial \theta} \frac{\partial \tilde f}{\partial \rho}
\end{equation}
with the Jacobian of the transformation,
\begin{equation}
\label{eq:Jac}
\frac{1}{\sqrt{g}} = \frac{B^2}{\varepsilon^2 (J+ \rlap{-} \iota I)}.
\end{equation}

Equations~\ref{pressure} and~\ref{velthermal} introduce the parallel momentum response of the thermal plasma. These are required for coupling to the geodesic acoustic waves, accounting for the geodesic compressibility in the frequency range of the geodesic acoustic mode (GAM) \cite{40,41}. The coupling between the equations of the EP and thermal plasma is done in the equation of the perturbation of the toroidal component of the vorticity (eq. 3), particularly through the fifth term on the right side, introducing the EP destabilizing effect caused by the gradient of the fluctuating EP density.

Equilibrium flux coordinates $(\rho, \theta, \zeta)$ are used. Here, $\rho$ is a generalized radial coordinate proportional to the square root of the toroidal flux function, and normalized to the unity at the edge. The flux coordinates used in the code are those described by Boozer \cite{42}, and $\sqrt g$ is the Jacobian of the coordinate transformation. All functions have equilibrium and perturbation components represented as: $ A = A_{eq} + \tilde{A} $.

The FAR3D code uses finite differences in the radial direction and Fourier expansions in the two angular variables. The numerical scheme is semi-implicit in the linear terms.

The finite Larmor radius and the electron-ion Landau damping effects are excluded from the simulations for simplicity. A preliminary parametric study identified an EP model that reproduce is, in the first approximation, a resonance with similar stability properties as the EIC in the experiment \cite{11}.

The present model was already used to study the AE activity in LHD \cite{43,44}, TJ-II \cite{45,46,47} and DIII-D \cite{48,49,50,51}, showing a reasonable agreement with the observations.

\subsection{Equilibrium properties}

A fixed boundary equilibrium from the VMEC code \cite{37} was calculated during the LHD shot 116190 after the destabilization of an EIC event including the EP component of the total pressure. This equilibria is used as the reference case in the different parametric studies. Such an assumption implies that the structure of the equilibrium pressure profile is the same for all the cases, and the force balance is well approximated, in first order approximation, scaling the reference model for the range of thermal $\beta$ values tested, between $0.1 \%$ to $2 \%$. It should be noted that the largest thermal $\beta$ considered is $2 \%$ because simulations with a higher value of the thermal $\beta$ require the recalculation of the equilibria, modified due to the outward displacement of the magnetic axis (Shafranov shift), so the force balance and the equilibria pressure profile deviate with respect to the reference case. Also, an LHD plasma with a thermal $\beta$ below $0.1 \%$ cannot be sustained. The electron density and temperature profiles were reconstructed by Thomson scattering data and electron cyclotron emission. Table~\ref{Table:1} shows the main parameters of the thermal plasma and table~\ref{Table:2} the details of the helically trapped EP population driven by the perpendicular hydrogen NBI in the reference model. The cyclotron frequency is $\omega_{cy} = 2.41 \cdot 10^{8}$ s$^{-1}$.

\begin{table}[t]
\centering
\begin{tabular}{c c c c}
\hline
$T_{i}$ (keV) & $n_{i}$ ($10^{20}$ m$^{-3}$) & $\beta_{th}$ ($\%$) & $V_{A}$ ($10^{7}$ m/s) \\ \hline
2 & 0.25 & 0.32 & 1.1 \\
\end{tabular}
\caption{Thermal plasma properties in the reference model (values at the magnetic axis). The first column is the thermal ion temperature, the second column is the thermal ion density, the third column is the thermal $\beta$ and the fourth column is the Alfv\' en velocity.} \label{Table:1}
\end{table}

\begin{table}[t]
\centering
\begin{tabular}{c}
Perpendicular NBI EP \\ 
\end{tabular}

\begin{tabular}{c c c}
\hline
$T_{f}$ (keV) & $n_{f}$ ($10^{20}$ m$^{-3}$) & $\beta_{f}$ ($\%$)\\ \hline
28 & 0.019 & 0.35 \\ 
\end{tabular}
\caption{Properties of the EP driven by the perpendicular NBI in the reference model (values at the magnetic axis). First column is the EP temperature, the second column is the EP density and the third column is the EP $\beta$.} \label{Table:2}
\end{table}

\begin{figure}[h!]
\centering
\includegraphics[width=0.45\textwidth]{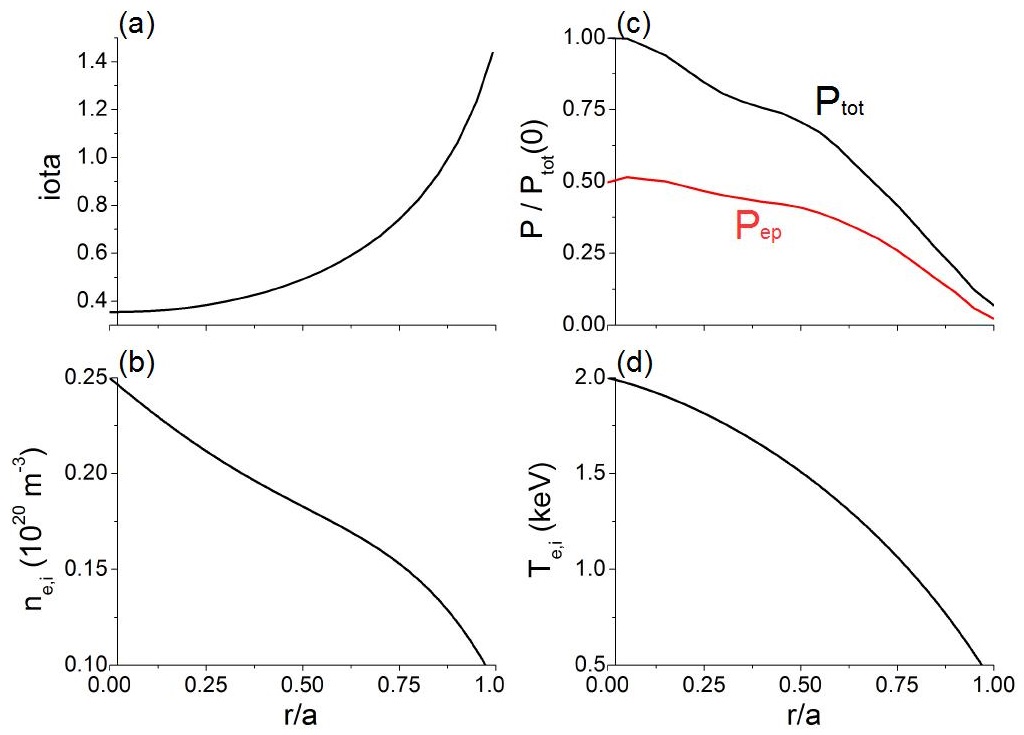}
\caption{(a) Iota profile, (b) thermal plasma density, (c) normalized pressure (thermal + EP pressure) and (d) thermal plasma temperature.}\label{FIG:1}
\end{figure}

The magnetic field at the magnetic axis is $2.5$ T and the averaged inverse aspect ratio is $\varepsilon=0.16$. The energy of the injected particles by the perpendicular NBI is $T_{f,\perp}(0) = 40$ keV, but we take the nominal energy $T_{f,\perp}(0) = 28$ keV ($v_{th,f} = 1.64 \cdot 10^{6}$ m/s) resulting in an averaged Maxwellian energy equal to the average energy of a slowing-down distribution. Figure~\ref{FIG:1} (a) shows the iota profile, (b) the thermal plasma density, (c) the normalized pressure (thermal plasma + EP pressure) and (d) the thermal plasma temperature. It should be noted that the effect of the equilibrium toroidal rotation is not included in the model for simplicity. The effect of the Doppler shift on the instability frequency caused by the toroidal rotation is small, particularly for a mode located in the plasma periphery, as it is observed in the experiments.

\subsection{Simulations parameters}

The dynamic and equilibrium toroidal (n) and poloidal (m) modes included in the study are summarized in table~\ref{Table:3}. The simulations are performed with a uniform radial grid of 1000 points. In the following, the mode number notation is $n/m$, which is consistent with the $\iota=n/m$ definition for the associated resonance.

\begin{table}[h]
\centering
\begin{tabular}{c || c | c}
\hline
n & 1 & 0  \\ \hline
m & $[1,2,3]$ & $[0,6]$ \\ \hline
\end{tabular}
\caption{Dynamic and equilibrium toroidal (n) and poloidal (m) modes in the simulations.} \label{Table:3}
\end{table}

The closure of the kinetic moment equations (6) and (7) breaks the MHD parities so both parities must be included for all the dynamic variables. Consequently, the different parities of a mode can show different growth rates and real frequencies in the eigenmode time series analysis. The convention of the code with respect to the Fourier decomposition is, in the case of the pressure eigenfunction, that $n > 0$ corresponds to $cos(m\theta + n\zeta)$ and $n<0$ corresponds to $sin(-m\theta - n\zeta)$. For example, the Fourier component for mode $-1/2$ is $\cos(-1\theta + 2\zeta)$ and for the mode $1/-2$ is $\sin(-1\theta + 2\zeta)$. The magnetic Lundquist number is assumed $S=5\cdot 10^6$.

\begin{figure}[h!]
\centering
\includegraphics[width=0.45\textwidth]{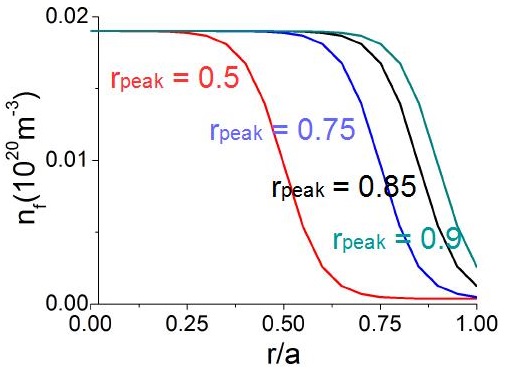}
\caption{EP density profile in the reference model ($r_{peak} = 0.85$) and other examples for different depositions regions of the perpendicular NBI.}\label{FIG:2}
\end{figure}

The density and temperature of the EP population generated by the perpendicular beam are calculated by the code MORH \cite{52,53}. For simplicity, no radial dependency of the EP energy is considered and the EP density profile given by MORH code is fitted to the following analytic expression:

\begin{equation}
\label{EP_dens}
$$n_{f,||}(r) = \frac{(0.5 (1+ \tanh(\delta_{r} \cdot (r_{peak}-r))+0.02)}{(0.5 (1+\tanh(\delta_{r} \cdot r_{peak}))+0.02)}$$
\end{equation}
with the location of the EP density gradient profile defined by the variable $r_{peak} = 0.85$ and the flatness by $\delta_{r} = 10$ in the reference model. Figure~\ref{FIG:2} shows the EP density profiles in the reference model and examples of other configurations where the perpendicular NBI is deposited in different plasma regions.

The ratio between the EP thermal velocity and the Alfv\' en velocity at the magnetic axis ($v_{th,f}/v_{A0}$) indicates the resonance coupling efficiency between the EP and the thermal plasma Alfven waves. This ratio is proportional to the square root of the EP temperature (NBI voltage) while the thermal plasma density is inverse proportional to the magnetic field magnitude. The Landau closure in the model is based on two moment equations for the energetic particles, which is equivalent to a two-pole approximation of the plasma dispersion relation. The closure coefficients are adjusted by fitting analytic AE growth rates. Such assumptions are consistent with a Lorentzian energy distribution function for the energetic particles. The lowest order Lorentzian can be matched either to a Maxwellian or to a slowing-down distribution by choosing an equivalent average energy. For the results given in this paper, we have matched the EP temperature to the mean energy of a slowing-down distribution function. 

\section{Stabilization trends observed experimentally \label{sec:exp}}

First, the LHD operation scenarios with stable $1/1$ EIC are identified with respect to the thermal plasma density and temperature. If the thermal plasma density is modified, the Alfven velocity also changes as well as the resonance between the helically trapped EP and bulk plasma. On the other hand, the variation of the thermal plasma temperature modifies the plasma resistivity. In both cases, the thermal plasma $\beta$ changes. 

\begin{figure}[h!]
\centering
\includegraphics[width=0.45\textwidth]{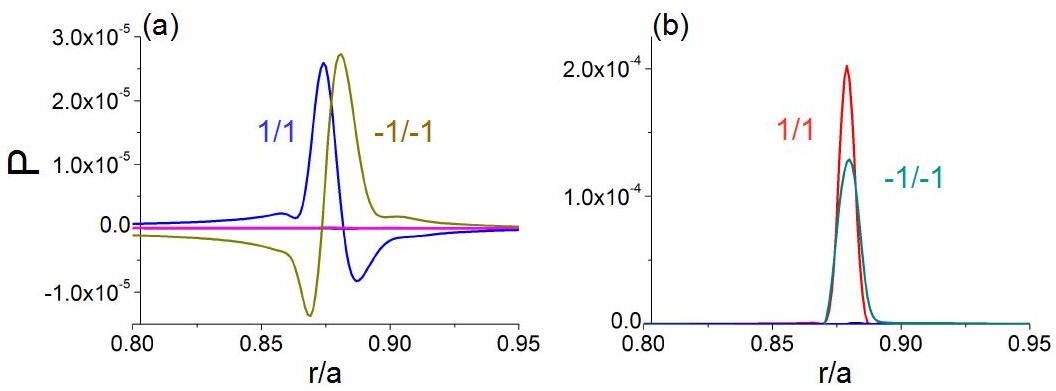}
\caption{(a) Eigenfunction of the $1/1$ EIC in the reference case. (b) Eigenfunction of the $1/1$ RIC if the destabilizing effect of the EP is not included in the reference case.}\label{FIG:3}
\end{figure}

The parametric scans are based upon the reference model where a $1/1$ EIC with a growth rate of $\gamma \tau_{A0} = 0.002$ and a frequency of $f = 4.8$ kHz is destabilized. Figure~\ref{FIG:3}a shows the pressure eigenfunction of the $1/1$ EIC that has a normalized width of $\Delta w_{p} / a = 0.05$ and the panel b a $1/1$ RIC with $\Delta w_{p} / a = 0.025$ unstable if the destabilizing effect of the EP is not included in the simulation.

\begin{figure}[h!]
\centering
\includegraphics[width=0.45\textwidth]{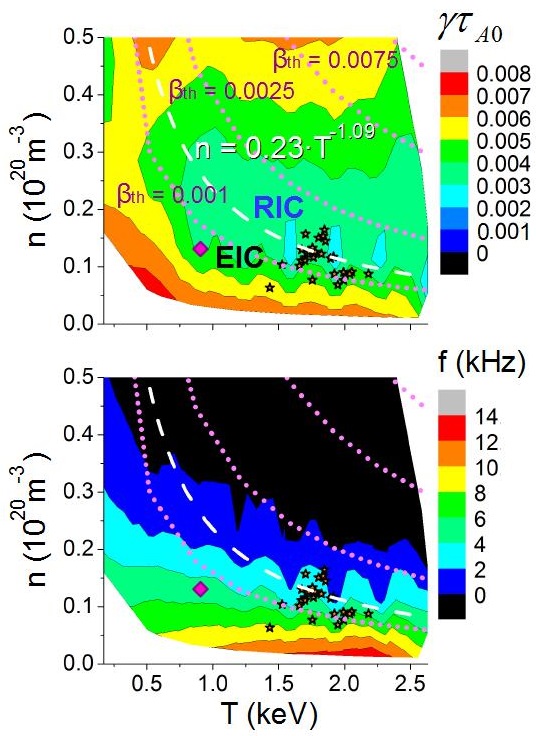}
\caption{Growth rate and frequency of the instabilities in simulations with different values of the thermal plasma density and temperature at the $\rlap{-} \iota = 1$ rational surface. The dashed purple lines indicate the iso-lines of the simulations with the same thermal $\beta$. The solid white line shows the transition between scenarios with dominant $1/1$ EIC and RIC fitted by the non-linear curve $n=aT^{b}$. The stars show the EIC destabilized in LHD discharges with respect to the thermal plasma density and temperature at the $\rlap{-} \iota = 1$ rational surface. The pink diamonds indicate the reference case.}\label{FIG:4}
\end{figure}

Figure~\ref{FIG:4} shows the growth rate (panel a) and frequency (panel b) of the instabilities calculated in simulations with different thermal plasma densities and temperatures at the $\rlap{-} \iota = 1$ rational surface.

The $1/1$ EIC are destabilized in LHD operational scenarios with low thermal plasma density and temperature. If the thermal plasma $\beta$ is above $0.25\%$ (dotted purple lines) the $1/1$ EIC are stable, although the EIC are unstable up to a thermal plasma $\beta$ of almost $0.2 \%$ if the thermal plasma density is below $0.15 \cdot 10^{20}$ m$^{-3}$ and the thermal plasma temperature is above $1.5$ keV. On the other hand, the EIC are also unstable to a thermal plasma density up to $0.25 \cdot 10^{20}$ m$^{-3}$ if the thermal plasma temperature is below $0.5$ keV. The dashed white line indicates the transition between LHD operation scenarios with unstable $1/1$ EIC and unstable RIC, fitted by a non-linear curved with $n=0.23T^{-1.09}$. The stars in the plots indicate $1/1$ EIC destabilized during several LHD discharges, showing a reasonable agreement with the $1/1$ EIC threshold predicted by the simulations. It should be noted that there are some $1/1$ EIC destabilized during LHD operation scenarios above the threshold predicted by the simulations, particularly for a thermal plasma temperature larger than $1.75$ keV. The disagreement can be explained by an increment of the EP $\beta$ associated with the increase of the thermal plasma temperature \cite{54}. On the other hand, the experimental data also shows a decrease of helically trapped EP as the thermal plasma density increases, so the EP $\beta$ decreases. Such effects are not included in the simulations because the variation of the EP $\beta$ with the thermal plasma density/temperature is known qualitatively but not quantitatively. Also, performing the simulations with a fixed EP $\beta$ helps to isolate the effect of the thermal plasma density and temperature on the $1/1$ EIC stability. Nevertheless, the effect of the EP $\beta$ on the $1/1$ EIC stability was analyzed in a previous study \cite{11}.

\begin{figure}[h!]
\centering
\includegraphics[width=0.45\textwidth]{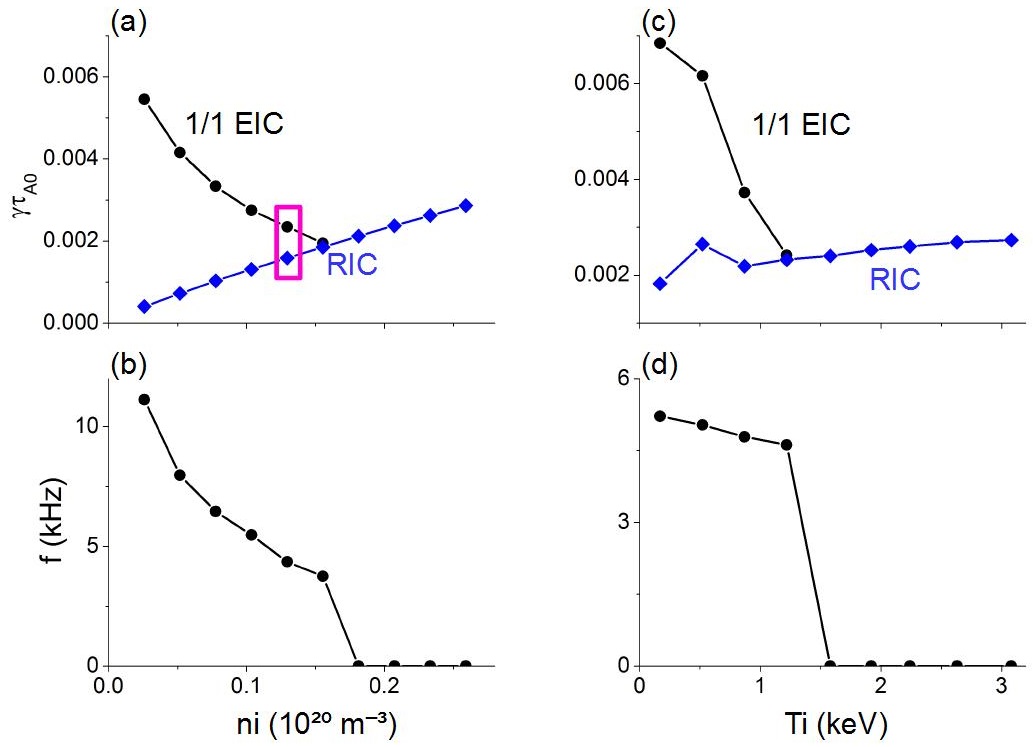}
\caption{Growth rate and frequency of the instabilities in parametric scans of the thermal plasma density (panels a and b) and temperature (panels c and d) at the $\rlap{-} \iota = 1$ rational surface. The black lines with black dots represent the $1/1$ EIC and the blue lines with blue diamonds the RIC. The pink rectangle indicates the reference case.}\label{FIG:5}
\end{figure}

To clarify the stabilization trends, a parametric scan is performed modifying individually the thermal plasma density and temperature at the $\rlap{-} \iota = 1$ rational surface, analyzing the growth rate and frequency of the $1/1$ EIC and RIC, shown in figure~\ref{FIG:5}.

If the thermal plasma density increases above $0.155 \cdot 10^{20}$ m$^{-3}$ at the $\rlap{-} \iota = 1$ rational surface the $1/1$ EIC are stabilized and the RIC are unstable (panels a and b). The decrease of the $1/1$ EIC growth rate as the thermal plasma density increases is caused by a weaker resonance between the helically trapped EP and the bulk plasma, due to a reduction of the Alfven velocity and an increase of the $v_{th,f}/v_{A0}$ ratio. On the other hand, the RIC growth rate is enhanced because the thermal plasma $\beta$ increases with the thermal plasma density, leading to a stronger destabilization of the pressure gradient driven modes. The parametric scan of the thermal plasma temperature shows the stabilization of the $1/1$ EIC above the $1.22$ keV at the $\rlap{-} \iota = 1$ rational surface (panels c and d). The $1/1$ EIC growth rate decreases as the plasma resistivity decreases, although the RIC growth rate increases due to the increase of the thermal plasma $\beta$. It should be noted that the further destabilization of the pressure gradient driven modes caused by a larger thermal $\beta$ dominates over the drive weakening caused by the decrease of the plasma resistivity. Consequently, the RIC growth rate enhancement with the thermal plasma temperature is weaker than the scaling with the thermal plasma density.

\begin{figure}[h!]
\centering
\includegraphics[width=0.45\textwidth]{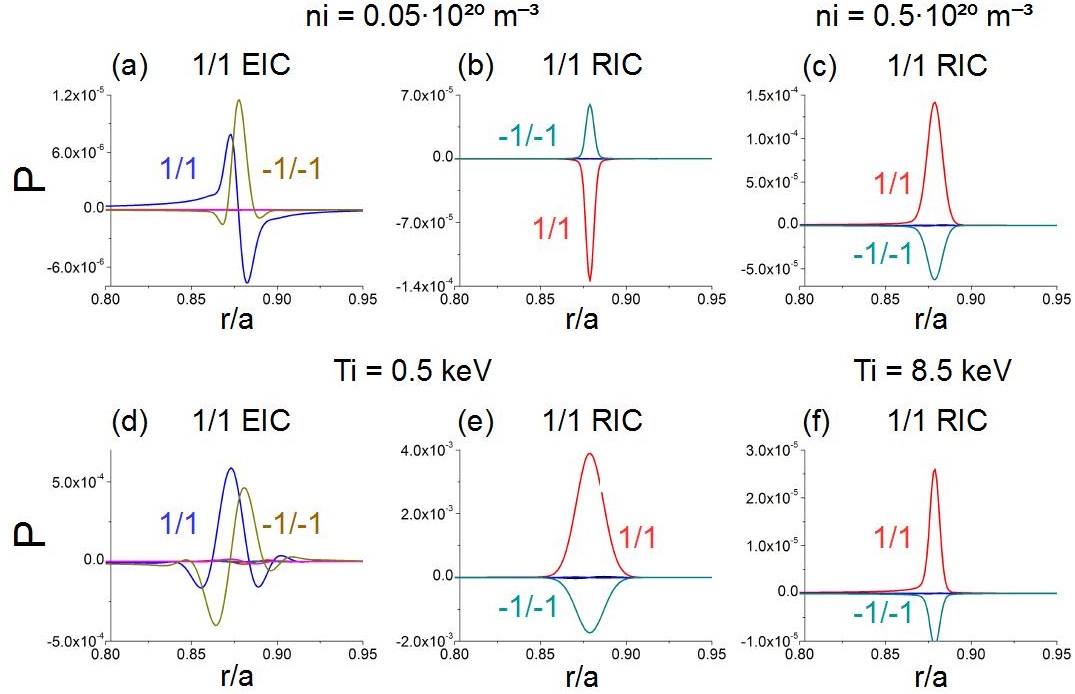}
\caption{Eigenfunction structure of the $1/1$ EIC and RIC if $n_{i} = 0.026 \cdot 10^{20}$ m$^{-3}$ (panels a and b) and $T_{i} = 0.17$ keV (panels d and e) at the $\rlap{-} \iota = 1$ rational surface. Eigenfunction structure of the RIC if $n_{i} = 0.26 \cdot 10^{20}$ m$^{-3}$ (panel c) and $T_{i} = 3.1$ keV (panel f)  at the $\rlap{-} \iota = 1$ rational surface.}\label{FIG:6}
\end{figure}

The effect of the thermal plasma $\beta$ and resistivity or the $1/1$ EIC and RIC stability can also be observed in the eigenfunction structure. Figure~\ref{FIG:6} shows the $1/1$ EIC and RIC eigenfunction structure in simulations with different thermal plasma densities and temperatures at the $\rlap{-} \iota = 1$ rational surface.

A decrease of the thermal plasma density (temperature) leads to a reduction (increase) of the normalized width of the $1/1$ EIC eigenfunction to $\Delta w_{p} / a = 0.04$ (0.07), see panels a and d. The trends are the same for the RIC: the normalized width of the eigenfunction increases from $\Delta w_{p} / a = 0.025$ to $0.032$ if the thermal plasma density increases (panels b and c), although it decreases from $\Delta w_{p} / a = 0.046$ to $0.02$ if the thermal plasma temperature increases (panels e and f).

In summary, the simulations reproduce the optimization trends observed experimentally regarding the $1/1$ EIC stability. Such trends can be explained by a weaker resonance between the helically trapped particle as the thermal plasma density increases and a reduction of the plasma resistivity as the thermal temperature increases. Nevertheless, an increment of the thermal plasma temperature also leads to a shorter slowing down time of the helically trapped particles before thermalization. Thus, this effect should also be considered in the analysis. Such a study is similar, in first order approximation, to analyzing the effect of the averaged thermalized velocity of the EP on the resonance properties. This study is performed in section \ref{sec:EP}.

\section{Magnetic field topology and intensity \label{sec:Bfield}}

\begin{figure}[h!]
\centering
\includegraphics[width=0.45\textwidth]{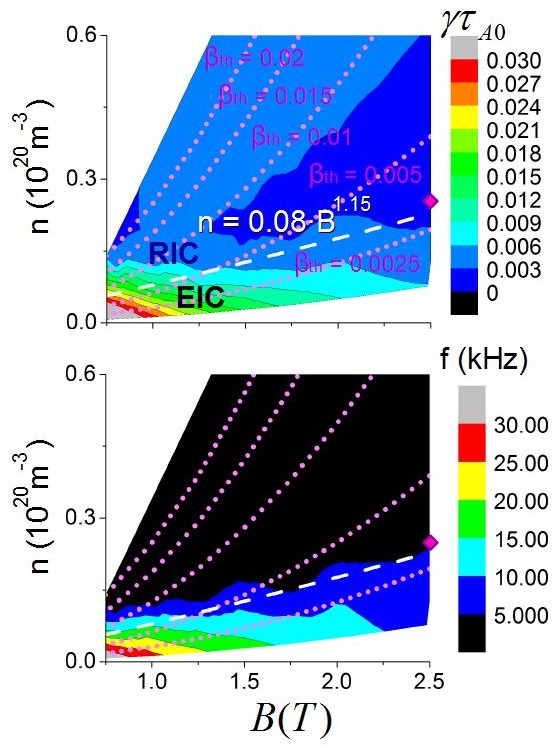}
\caption{Growth rate and frequency of the instabilities for different values of the thermal plasma density and magnetic field intensity at the magnetic axis. The dotted purple lines indicate the iso-lines of the simulation with the same thermal $\beta_{0}$ value. The dashed white line shows the transition between scenarios with dominant $1/1$ EIC to those with RIC. This can be fitted by the non-linear curve $n=aB^{b}$. The pink diamonds indicate the reference case.}\label{FIG:7}
\end{figure}

The optimization trends related to the magnetic field magnitude and rotational transform are analyzed in this section. If the magnetic field intensity is modified, the thermal and EP $\beta$ also change, because $\beta$ is inversely proportional to the square of the magnetic field magnitude. In addition, the resonance between the helically trapped EP and the bulk plasma is altered because the Alfven velocity is proportional to the magnetic field magnitude, thus the $v_{th,f}/v_{A0}$ ratio changes, as well as the plasma cyclotron frequency. Consequently, the stability properties of the $1/1$ EIC and RIC are modified if the LHD magnetic field intensity changes. Figure~\ref{FIG:7} shows the instabilities growth rate (panel a) and frequency (panel b) in LHD operation scenarios with different magnetic field intensity and thermal plasma densities (fixed the thermal plasma temperature to $2$ keV).

\begin{figure}[h!]
\centering
\includegraphics[width=0.45\textwidth]{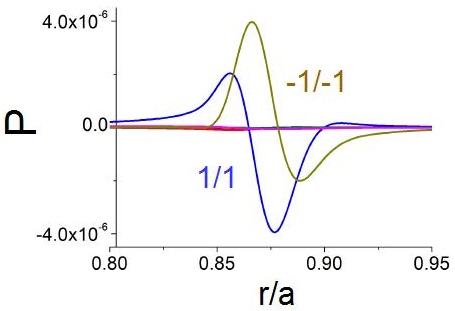}
\caption{Eigenfunction of $1/1$ EIC in an LHD operational scenario with $B=0.75$ T.}\label{FIG:8}
\end{figure}

The growth rate of the $1/1$ EIC is enhanced if the magnetic field intensity and the thermal plasma density at the magnetic axis decrease with respect to the reference case. The dashed white line indicates the transition between LHD operation scenarios with unstable $1/1$ EIC to those with unstable RIC, fitted by a non-linear curved with $n=0.08B^{1.25}$. The LHD operational scenarios with a magnetic field intensity above $1.5$ T and a thermal $\beta_{0}$ below $0.5 \%$ show unstable $1/1$ EIC, although in operation scenarios with lower magnetic field intensity the $1/1$ EIC can be destabilized for a thermal $\beta_{0}$ up to $0.5 \%$. In addition, the $1/1$ EIC frequency increases up to $35$ kHz as the magnetic field intensity and the thermal plasma density decrease. Consequently, the $1/1$ EIC are easily destabilized in LHD operation scenarios with a low magnetic field intensity, although if the thermal plasma density at the magnetic axis is above $0.1 \cdot 10^{20}$ m$^{-3}$ ($\beta_{0} > 0.5 \%$) the $1/1$ EIC are stable. Figure~\ref{FIG:8} shows the $1/1$ EIC eigenfunction in a simulation with a magnetic field intensity of $0.75$ T, indicating an increase of the eigenfunction normalized width up to $\Delta w_{p} / a = 0.08$.

\begin{figure}[h!]
\centering
\includegraphics[width=0.45\textwidth]{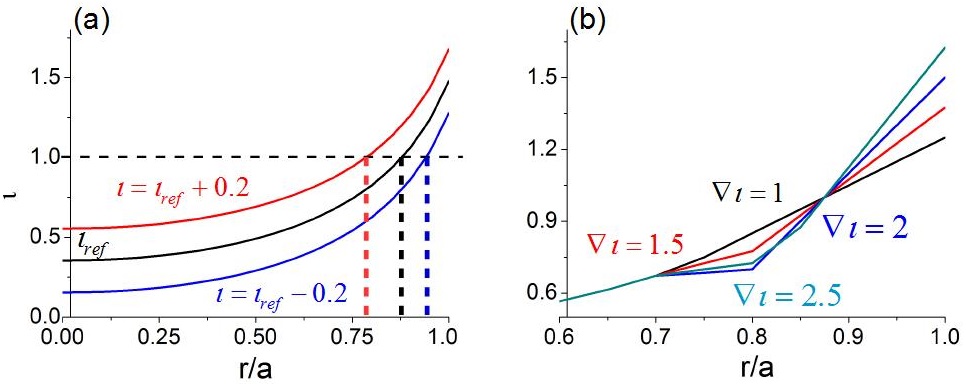}
\caption{Iota profile in the simulations where the location of the $\rlap{-} \iota = 1$ rational surface along the normalized minor radius (panel a) or the magnetic shear in the plasma periphery (panel b) are modified.}\label{FIG:9}
\end{figure}

The effect of the rotational transform on the $1/1$ EIC stability is also analyzed. First, the effect of the location of the $\rlap{-} \iota = 1$ rational surface along the normalized minor radius is studied. This analysis is performed displacing the iota profile by $\rlap{-} \iota = \rlap{-} \iota_{ref} + \Delta \rlap{-} \iota$ with $\Delta \rlap{-} \iota = 0.02$ between $[-0.2 , 0.2]$, see figure~\ref{FIG:9}a. Second, the effect of the magnetic shear in the plasma periphery is analyzed increasing the iota profile slope around $\rlap{-} \iota = 1$ rational surface from $d \rlap{-} \iota  / d \rho = 1$ to $2.5$, see figure~\ref{FIG:9}b.

\begin{figure}[h!]
\centering
\includegraphics[width=0.45\textwidth]{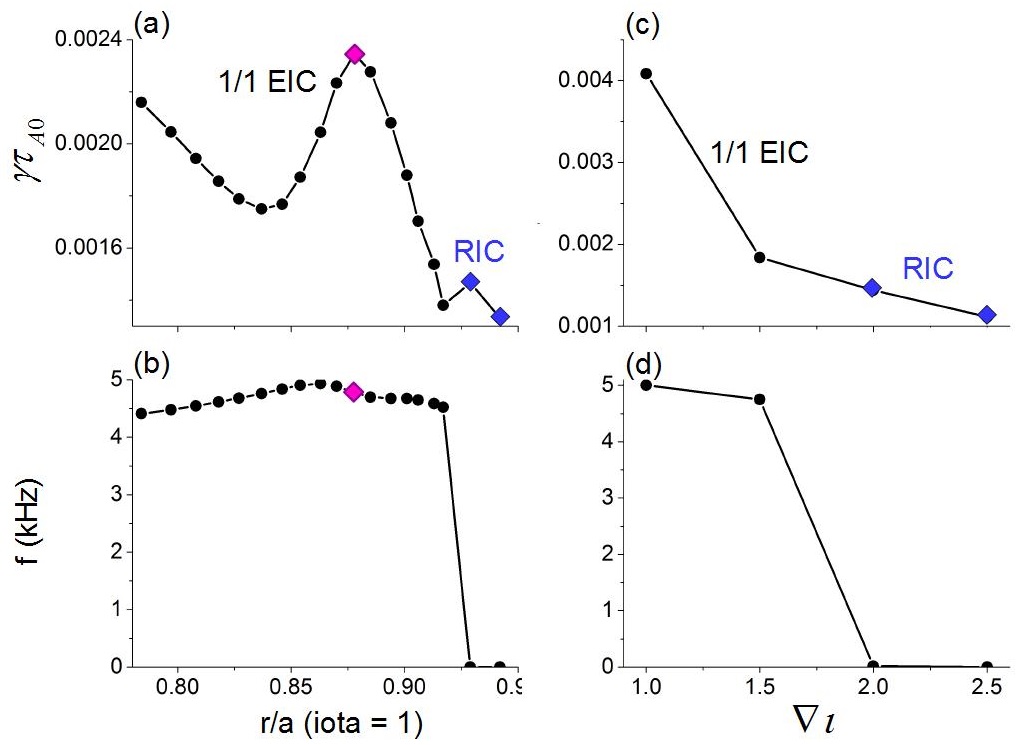}
\caption{Growth rate and frequency of the instability if the location of the $\rlap{-} \iota = 1$ rational surface along the normalized minor radius (panels a and b) or the magnetic shear (panels c and d) are modified. The black dots indicate the destabilization of $1/1$ EIC and the blue diamonds $1/1$ RIC. The pink diamonds indicate the reference case.}\label{FIG:10}
\end{figure}

Figure~\ref{FIG:10} shows the growth rate and frequency of the instability with respect to the location of the $\rlap{-} \iota = 1$ rational surface along the normalized minor radius (panels a and b) or the magnetic shear (panels c and d). 

The $1/1$ EIC are stabilized if the $\rlap{-} \iota = 1$ rational surface is located further outward from $r/a = 0.925$. The largest growth rate is reached in the simulation with the $\rlap{-} \iota = 1$ rational surface located at $r/a = 0.878$ ($0.87$ in the reference case). This is caused by a strong resonance between the helically trapped EP and bulk plasma, because the gradient of the EP density profile is located closer to the $\rlap{-} \iota = 1$ rational surface, which determines the source of free energy to destabilize the $1/1$ EIC. On the other hand, the $1/1$ EIC growth rate increases again if the $\rlap{-} \iota = 1$ rational surface is located inward from $r/a < 0.818$, because in this plasma region the magnetic shear is weaker. Following the analysis of the magnetic shear effects, the threshold to stabilize the $1/1$ EIC is $d \rlap{-} \iota / d \rho = 2.0$. 

\begin{figure}[h!]
\centering
\includegraphics[width=0.45\textwidth]{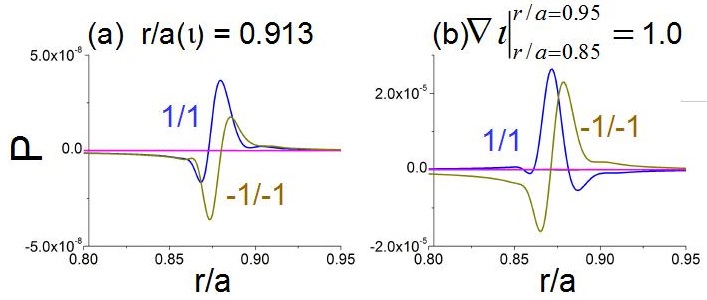}
\caption{Eigenfunction of the $1/1$ EIC if the $\rlap{-} \iota = 1$ rational surface is located at $r/a = 0.913$ (panel a). Eigenfunction of the $1/1$ EIC if $d \rlap{-} \iota / d \rho = 1.0$ (panel b)}\label{FIG:11}
\end{figure}

Figure~\ref{FIG:11} shows the $1/1$ EIC eigenfunction for different locations of the $\rlap{-} \iota = 1$ rational surface (panels a and b) and magnetic shear (panel c).

The normalized width of the $1/1$ EIC eigenfunction decreases compared with the reference case if the $\rlap{-} \iota = 1$ rational surface is located at $r/a = 0.913$ (panel a, $\Delta \omega_{p} / a = 0.0375$), because the $\rlap{-} \iota = 1$ rational surface is located in a plasma region with stronger magnetic shear and away from the gradient of the EP density profile. On the other hand, for a simulation with a weaker magnetic shear than the reference case (panel b), $d \rlap{-} \iota / d \rho = 1.0$, the width of the eigenfunction increases up to $\Delta \omega_{p} / a = 0.055$.

Consequently, optimization trends to stabilize the $1/1$ EIC are identified for LHD operational scenarios with a strong magnetic field intensity, a magnetic field topology with the $\rlap{-} \iota = 1$ rational surface located further outward from $r/a = 0.9$ and a strong magnetic shear in the plasma periphery. 

\section{Operational regime of the perpendicular NBI \label{sec:EP}}

In this section the $1/1$ EIC stability is analyzed with respect to the operational regime of the perpendicular NBI, particularly considering variations in the voltage and the deposition region. If the NBI voltage increases the EP temperature is higher, modifying the averaged thermalized velocity of the EP, proportional to the square root of the EP temperature. Consequently, the resonance between the helically trapped EP and the bulk plasma also changes. On the other hand, varying the deposition region of the perpendicular NBI changes the location of the EP density gradient along the normalized minor radius.

\begin{figure}[h!]
\centering
\includegraphics[width=0.45\textwidth]{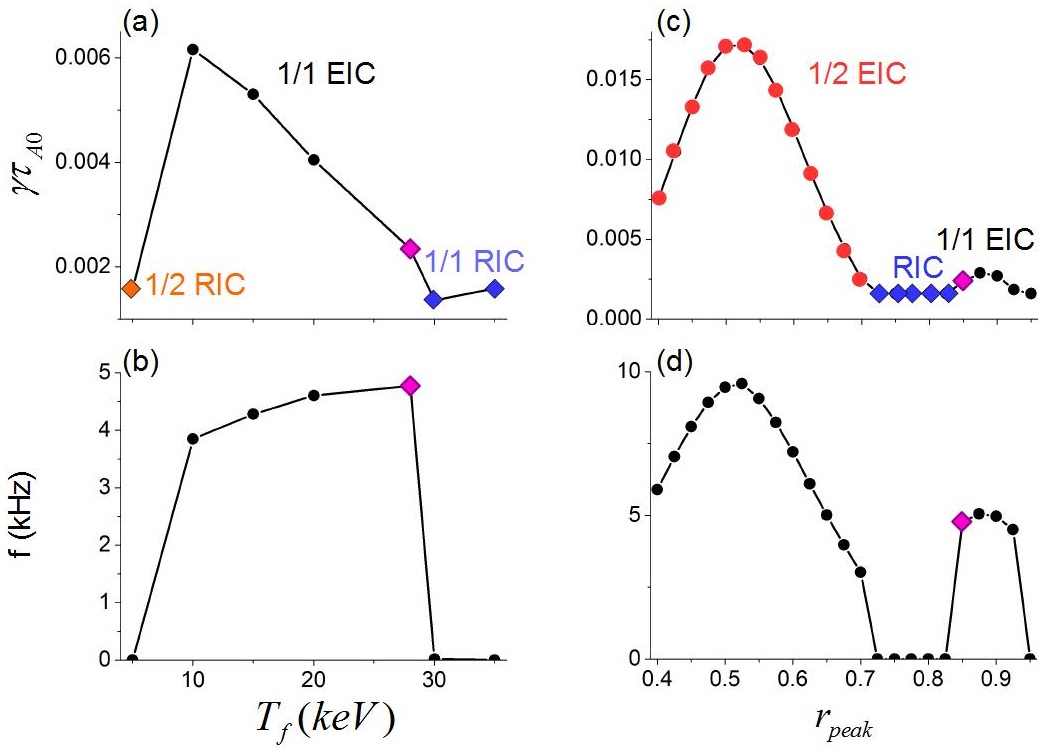}
\caption{Growth rate and frequency of the instability if the NBI voltage (panels a and b) or the deposition region (panels c and d) are modified. The black dots indicate the destabilization of $1/1$ EIC, the red dots $1/2$ EIC, the blue diamonds $1/1$ RIC and the orange diamonds $1/2$ RIC. The pink diamonds indicate the reference case.}\label{FIG:12}
\end{figure}

Figure~\ref{FIG:12} shows the growth rate and frequency of the instability if the NBI voltage (panels a and b) and the NBI deposition region (panels c and d) are modified. It should be noted that  the EP $\beta$ is fixed in the simulations, so the decrease/increase of the EP energy is compensated by an increase/decrease of the EP density.

Decreasing the voltage of the perpendicular NBI, and hence the EP temperature, the growth rate of the $1/1$ EIC is enhanced reaching a local maximum for $T_{f} = 10$ keV (panels a and b). On the other hand, if the EP temperature is reduced to $5$ keV the $1/1$ EIC are stable and a $1/2$ RIC is destabilized in the middle plasma region. In addition, if the EP temperature is $T_{f} \ge 30$ keV, the $1/1$ EIC are stabilized and $1/1$ RIC are unstable. Consequently, following the discussion initiated in section~\ref{sec:exp}, the LHD operation scenarios with a high thermal plasma temperature have a smaller slowing down time of the EP, thus the averaged thermalized velocity of the EP is higher, leading to the stabilization of the $1/1$ EIC because the resonance of the EP and bulk plasma changes. The effect of the NBI deposition region is examined in panels c and d. If the gradient of the EP density profile is located between $r/a = 0.725 - 0.825$ the $1/1$ EIC are stable and the RIC are destabilized. On the other hand, if the NBI is deposited further inward, between $r/a = 0.4 - 0.7$, $1/2$ EIC are destabilized. Similarly, if the NBI is deposited outward with respect to the reference case, the local maximum of the $1/1$ EIC growth rate is reached at $r/a = 0.875$, decreasing from $r/a = 0.9$. It should be noted that the tilt of the perpendicular NBI cannot be modified in LHD to change the deposition region, although the deposition region varies if the vacuum magnetic axis ($R_{ax}$) is displaced inward (LHD inward shifted configuration with $R_{ax} < 3.6$ m) or outward (LHD outward shifted configuration with $R_{ax} > 3.7$ m). Inward shifted configurations may lead to an NBI deposition region located slightly outward with respect to the reference case (default LHD configuration with $R_{ax} = 3.6$ m), so the growth rate of the $1/1$ EIC should increase, although above a given $R_{ax}$ threshold the growth rate decreases. Likewise, outward shifted configurations may lead to an NBI deposition region located slightly inward, leading to weaker $1/1$ EIC and the stabilization below a $R_{ax}$ threshold.

\begin{figure}[h!]
\centering
\includegraphics[width=0.45\textwidth]{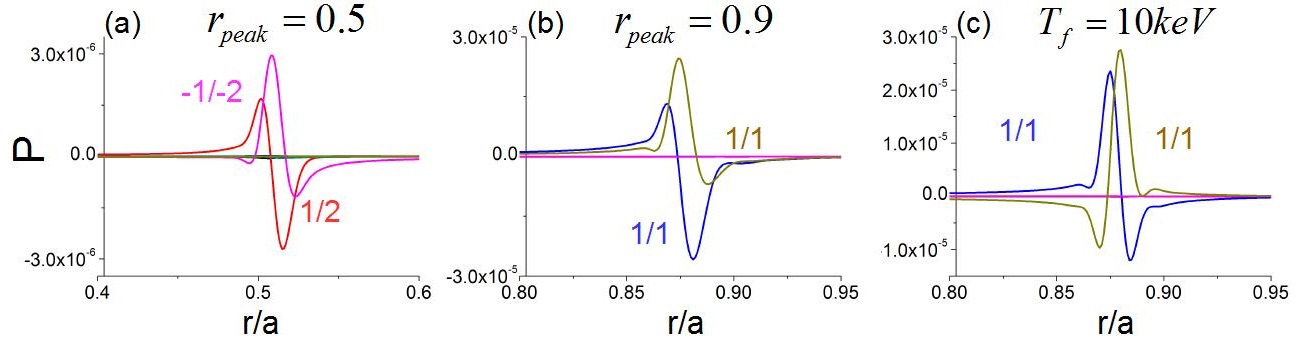}
\caption{(a) Eigenfunction of the $1/2$ EIC if the EP profile density gradient is located at $r/a = 0.5$. (b) Eigenfunction of the $1/1$ EIC if the EP profile density gradient is located at $r/a = 0.9$. (c) Eigenfunction of the $1/1$ EIC if the EP temperature is $T_{f} = 10$ keV.}\label{FIG:13}
\end{figure}

Figure~\ref{FIG:13} shows the $1/2$ EIC eigenfunction, destabilized if the gradient of the EP density profile is located at $r/a = 0.5$ (panel a), the $1/1$ EIC eigenfunction if the EP profile density gradient is located at $r/a = 0.9$ (panel b) or the EP temperature is $T_{f} = 10$ keV (panel c). The normalized width of the $1/1$ EIC eigenfunction ($\Delta w_{p} / a = 0.035$) decreases if the NBI is deposited at $r/a=0.9$. On the other hand, the width of the $1/1$ EIC eigenfunction ($\Delta w_{p} / a = 0.0525$) increases if the NBI voltage decreases.

The simulation results indicate optimization trends to stabilize the $1/1$ EIC if the perpendicular NBI voltage increases, leading to a weakly or non resonant NBI operation regime. This has already been observed in the analysis of the TAE stability in LHD and for the HAE in TJ-II \cite{44,47}. In addition, if the NBI is deposited more inward, the $1/1$ EIC can be stabilized. This occurs for the case of outward shifted LHD configurations.

\section{Thermal plasma and perpendicular NBI species \label{sec:Species}}

In this section the stability of the $1/1$ EIC is analyzed for Hydrogen, Deuterium, Hydrogen+Helium and Deuterium+Helium plasma heated by a perpendicular beam injecting Hydrogen or Deuterium. Modifying the thermal plasma and perpendicular NBI species, particularly the atomic number, the Alfven velocity changes leading to a shift in the resonance between the EP and thermal plasma.

\begin{figure}[h!]
\centering
\includegraphics[width=0.45\textwidth]{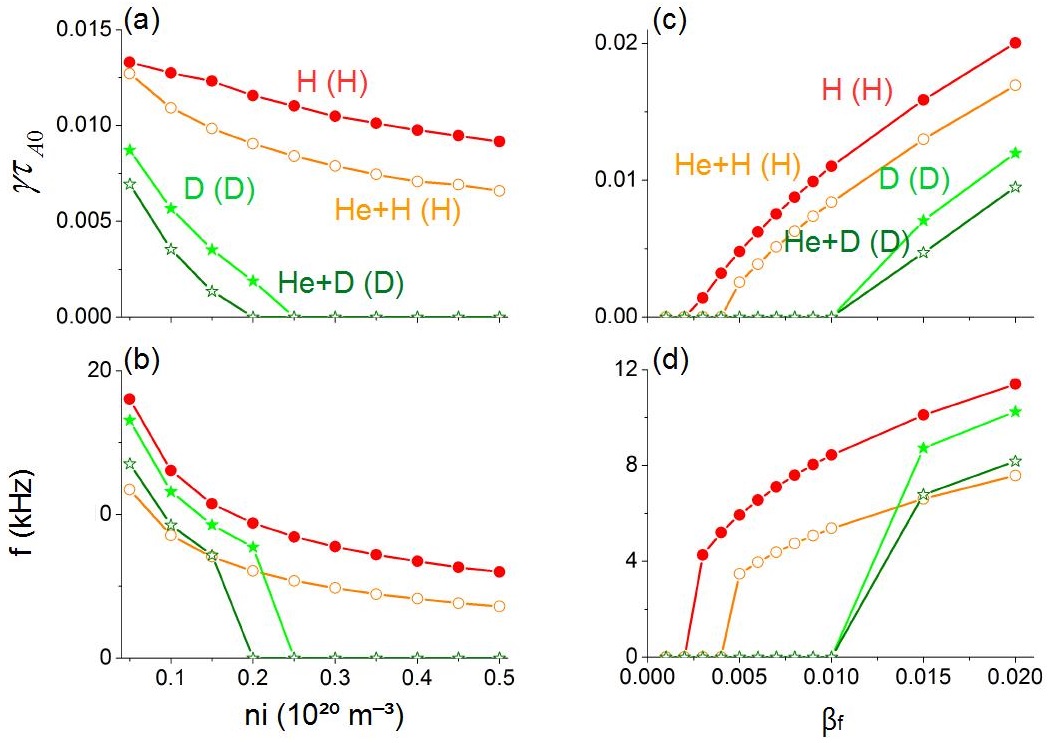}
\caption{Growth rate (panel a) and frequency (panel b) of the $1/1$ EIC for different thermal plasma densities ($\beta_{f}=0.01$) in Hydrogen plasma with Hydrogen NBI (red line and dots), Helium + Hydrogen plasma with Hydrogen NBI (orange line and circles), Deuterium plasma with Deuterium NBI (green line and stars) and Deuterium + Helium plasma with Deuterium NBI (dark green line and stars). Growth rate (panel c) and frequency (panel d) of the $1/1$ EIC for different EP $\beta$ (fixed $n_{i} = 0.25 \cdot 10^{20}$ m$^{-3}$).}\label{FIG:14}
\end{figure}

Figure~\ref{FIG:14}, panels a and b, shows the growth rate and frequency of the $1/1$ EIC for different thermal plasma, NBI species and thermal plasma densities (fixed $\beta_{f} = 1 \%$). The thermal plasma density threshold to stabilize the $1/1$ EIC in a Deuterium plasma is $n_{i} = 0.25 \cdot 10^{20}$ m$^{-3}$, smaller compared to a Hydrogen plasma where the $1/1$ EIC are not stabilized for thermal plasma density up to $n_{i} = 0.5 \cdot 10^{20}$ m$^{-3}$. In addition, if a Hydrogen or Deuterium plasma is mixed with Helium, the growth rate of the $1/1$ EIC is smaller for all the thermal plasma densities tested and the stabilization threshold decreases, $n_{i} = 0.2 \cdot 10^{20}$ m$^{-3}$ for a Deuterium + Helium plasma. Also, the $1/1$ EIC frequency decreases in a Deuterium plasma compared with a Hydrogen plasma, just as for a Hydrogen or Deuterium plasma mixed with Helium. To confirm the optimization trend, the growth rate and frequency of the $1/1$ EIC are analyzed for different EP $\beta$ if the thermal plasma density is fixed to $0.25 \cdot 10^{20}$ m$^{-3}$, see panels c and d. The $1/1$ EIC destabilization threshold is $\beta_{f} = 1\%$ in a Deuterium plasma and $\beta_{f} = 0.3\%$ in a Hydrogen plasma. In addition, the growth rate of the $1/1$ EIC further reduces if Helium is added to the thermal plasma.

\begin{figure}[h!]
\centering
\includegraphics[width=0.45\textwidth]{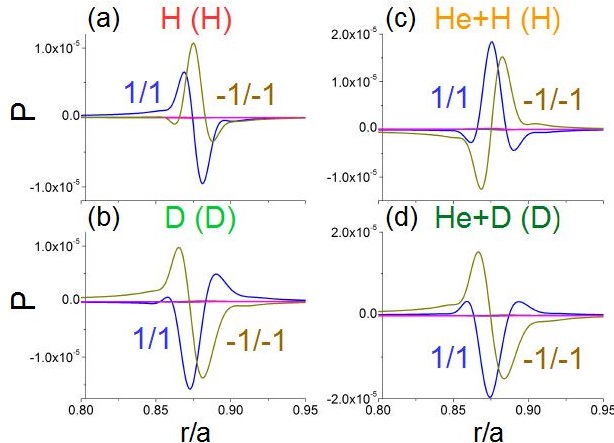}
\caption{Eigenfunction of the $1/1$ EIC for a Hydrogen plasma heated by a Hydrogen NBI (a), a Deuterium plasma heated by a Deuterium NBI (b), a Hydrogen + Helium plasma heated by a Hydrogen NBI (c) and a Deuterium + Helium plasma heated by a Deuterium NBI (d) for $\beta_{f} = 1.5 \%$.}\label{FIG:15}
\end{figure}

Figure~\ref{FIG:15} shows the $1/1$ EIC eigenfunction for a Hydrogen plasma heated by a Hydrogen NBI (panel a), a Deuterium plasma heated by a Deuterium NBI (panel b), a Hydrogen + Helium plasma heated by a Hydrogen NBI (panel c) and a Deuterium + Helium plasma heated by a Deuterium NBI (panel d) for a EP $\beta$ of $1.5 \%$. The normalized width of the $1/1$ EIC eigenfunction is larger in a Deuterium plasma with respect to a Hydrogen plasma or a plasma mixed with Helium.

From fig~\ref{FIG:14}, a Deuterium plasma shows an improved $1/1$ EIC stability compared to a Hydrogen plasma; this is further improved if the Deuterium is mixed with Helium. This is caused by a change in the $v_{th,f}/v_{A0}$ ratio, that is to say, resonance coupling efficiency between the EP and the thermal plasma Alfven waves. The simulation results are consistent with the experimental observations showing a higher thermal plasma density and perpendicular NBI injection intensity threshold to destabilize $1/1$ EIC in Deuterium plasma with respect to Hydrogen plasma \cite{27}.

\section{Multiple EP components \label{sec:multiple}}

The effect of the multiple EP species can modify the growth rate and frequency of the AEs and EPMs as it was observed in the TFTR experiment \cite{55,56,57}, as well as in theoretical studies that analyzed the stabilizing effect of the NBI driven EP on the AEs caused by $\alpha$ particle in ITER plasma and the AE stability in LHD /DIII-D plasma heated by multiple NBI lines \cite{58,59}. In the case of LHD discharges with EIC events, two different EP components coexist: the passing EP particles driven by the tangential NBI and the helically trapped EP driven by the perpendicular NBI. A previous theoretical study analyzed the effect of the tangential NBI injection intensity on the EIC stability, indicating that the EIC growth rate decreases as the tangential NBI injection intensity increases \cite{11}. The aim of this section is to identify whether this optimization trend is observed in the experimental data. To that end, the time interval between EIC events ($\Delta$) is studied with respect to the injection intensity of the perpendicular and tangential NBI. If the time interval between EIC increases, the EIC are less unstable.

\begin{figure}[h!]
\centering
\includegraphics[width=0.45\textwidth]{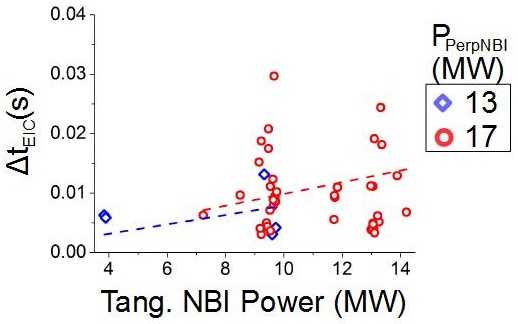}
\caption{Time interval between EIC events for different tangential NBI injection intensities for a fixed value of the perpendicular NBI injection intensity: blue diamonds show the discharges with a $P_{P,NBI} = 13.5 \pm 0.5$ MW and the red circles $P_{P,NBI} = 17.5 \pm 0.5$ MW. The plot includes the linear fit $\Delta t = b P_{T,NBI}$ for a perpendicular NBI injection of $13.5 \pm 0.5$ MW (dashed blue line) and $17.5 \pm 0.5$ MW (dashed red line)}\label{FIG:16}
\end{figure}

Fig~\ref{FIG:16} shows shows the EIC $\Delta$t for different tangential NBI injection intensities at a fixed value of the perpendicular NBI injection power: $P_{P,NBI} = 13.5 \pm 0.5$ MW (blue diamonds) and $P_{P,NBI} = 17.5 \pm 0.5$ MW (red circles). The study is limited to discharges with strong perpendicular NBI power $P_{P,NBI} > 13$ MW, because the available data is larger with respect to discharges with lower $P_{P,NBI}$ and the trends are easily identified due to the stronger destabilizing effect of the perpendicular NBI.  There is a trend that indicates a larger time spacing of the EIC as the tangential NBI power increases.  The linear regressions of the experimental data show an increase of the EIC $\Delta$ with the tangential NBI injection intensity: $\Delta t = 0.001 \cdot P_{T,NBI}$ ($17.5$ MW case) and $\Delta t = 0.0008 \cdot P_{T,NBI}$ ($13.5$ MW case). It should be noted that the verification of this optimization trend requires a larger data base.

\section{Conclusions and discussion \label{sec:conclusions}}

A set of linear simulations have been performed by the FAR3d code reproducing the optimization strategies explored in LHD experiments with respect to the thermal plasma parameters for stabilizing the $1/1$ EIC. The simulation results are in a reasonable agreement with the experimental data. 

The analysis identified the LHD operation scenario for different thermal $\beta$ values at the $\rlap{-} \iota = 1$ rational surface (fixed the magnetic field intensity) with unstable $1/1$ EIC. The study shows stable $1/1$ EIC if the thermal plasma density at the $\rlap{-} \iota = 1$ rational surface is above a certain threshold. The stabilization is caused by the decrease of the Alfven velocity and, thereby, a weaker resonance between the helically trapped EP and the bulk plasma. The stabilization of the $1/1$ EIC above a given threshold of the thermal plasma temperature is partly caused by a decrease of the plasma resistivity, leading to a narrow eigenfunction width. In addition, a higher thermal plasma temperature results in a decreased slowing down time of the EP, leading to a larger averaged thermalized EP velocity, also weakening the resonance. Thus, the application of electron cyclotron heating to increase the thermal plasma temperature near the $\rlap{-} \iota = 1$ rational surface, or increasing the thermal plasma density in the plasma periphery by controlled gas puffing, are methodologies that can stabilize the $1/1$ EIC and improve the LHD performance.

The threshold identified by the numerical model for the transition between the $1/1$ EIC and RIC with respect to the thermal plasma density/temperature at the $\rlap{-} \iota = 1$ rational surface is compared with LHD discharges where the $1/1$ EIC are destabilized. There is a reasonable agreement between the simulations and experimental data, although some $1/1$ EIC events are destabilized above the theoretical threshold during discharges with a thermal plasma temperature above $1.75$ keV. This disagreement can be explained by an increase of the EP $\beta$ with the thermal plasma temperature not included in the simulations.

Other optimization trends are analyzed to stabilize the $1/1$ EIC with respect to the LHD magnetic field intensity and topology. LHD operation scenarios with a low magnetic field, particularly if the $B=0.75$ T, could lead to the destabilization of $1/1$ EIC with high growth rates and frequencies up to $35$ kHz, thus this LHD regime should be avoided. On the other hand, the outward displacement of the $\rlap{-} \iota = 1$ rational surface leads to a decrease of the $1/1$ EIC growth rate, and stability if the rational surface is located further away from $r/a = 0.9$. In addition, the $1/1$ EIC are stabilized by the effect of the magnetic shear above a given threshold. Consequently, the optimization trends indicate that the $\rlap{-} \iota = 1$ rational surface must be located in a plasma region with large magnetic shear and away from the perpendicular NBI deposition region. Local variations of the rotational transform are possible thanks to the application of the electron cyclotron current drive (ECCD) on the plasma periphery, although this approach is limited by the flexibility of the ECH antenna system on LHD. The current drive efficiency of this system depends strongly on the magnetic field intensity (optimal for $B=1.375$ T).

A careful selection of the operational regime of the perpendicular NBI, particularly the voltage and the deposition region, can also stabilize the $1/1$ EIC. The simulations indicate that a NBI deposition located between $r/a = 0.725 - 0.825$ can stabilize the $1/1$ EIC. On the other hand, if the NBI deposition region is located further inward, between $r/a = 0.4 - 0.7$, the $1/2$ EIC can be destabilized showing a local maximum of the growth rate $6$ times larger than the $1/1$ EIC for a deposition region around $r/a = 0.5$. Nevertheless, the $1/1$ EIC growth rate decreases even for small displacements of the NBI deposition region with respect to the location of the $\rlap{-} \iota = 1$ rational surface along the normalized minor radius. Such an optimization trend can be explored experimentally if the LHD vacuum magnetic axis is displaced inward (inward shiflted LHD configurations with $R_{ax} < 3.6$ m) or outward (outward shiflted LHD configurations with $R_{ax} > 3.7$ m), because the tilt of the perpendicular NBI in LHD is fixed. Another optimization trend identified is the stabilization of the $1/1$ EIC if the NBI voltage increases, leading to a non resonant NBI operational regime if the EP temperature is above $30$ keV. Such threshold changes depending on the thermal plasma parameters. The threshold increases as the thermal plasma density and temperature or the magnetic field intensity decrease, because the $1/1$ EIC are destabilized with a weaker drive of the helically trapped EP. The voltage of the perpendicular NBI in LHD is fixed, thus this optimization trend can be only observed as a side effect of the thermal plasma temperature variation. In LHD operation scenarios with a high thermal plasma temperature, the EP slowing down time is smaller, thus the averaged EP thermal velocity is larger. This changes the EP resonance with the bulk plasma.

The analysis of the thermal plasma and NBI species on the $1/1$ EIC stability indicates that a Deuterium plasma shows a higher threshold for destabilizing the $1/1$ EIC with respect to the thermal plasma density and perpendicular NBI driving compared to a Hydrogen plasma. The $1/1$ EIC stability can be further improved if the Deuterium is mixed with Helium. This optimization trend was already confirmed in the LHD experiments \cite{27,60}.

The multiple EP species effects are also studied. The simplified statistical analysis performed with the experimental data shows an increment of the time interval between EICs as the power of the tangential NBI increases. Nevertheless, due to the reduced amount of available data, a larger database is required to confirm this optimization trend.

It should be noted that there are other successful approaches used to suppress the $1/1$ EIC not included in the present study, for example, the application of resonant magnetic perturbations near the $\rlap{-} \iota = 1$ rational surface. Such analysis cannot be done with the existing version of the FAR3d code because the numerical model is based on a VMEC equilibria that does not allow magnetic islands. 

The present study supports the optimization scenarios proposed previously by other authors \cite{28,29,30}, identifying the stabilization of the $1/1$ EIC as the thermal plasma density and temperature increase. The other optimization trends are identified with respect to the LHD magnetic field intensity and rotational transform as well as the perpendicular NBI operational regime must be verified experimentally. Dedicated LHD experiments will be performed in future LHD campaigns to study the $1/1$ EIC stability with respect to changes in the LHD magnetic field magnitude and vacuum magnetic axis location.

\ack

The authors would like to thank the LHD technical staff for their contributions in the operation and maintenance of LHD. This work was supported by NIFS07KLPH004.

\hfill \break

\end{document}